\documentclass[onecolumn]{article}    

\usepackage{graphicx}          
                               
\usepackage{amssymb}
\usepackage{amsthm}
\usepackage{gensymb}
\usepackage{subfig}
\usepackage{mathtools}
\usepackage{authblk}
\usepackage{hyperref}

\usepackage[a4paper, rmargin = 2.5cm, lmargin = 2.5cm]{geometry}

\newtheorem{theorem}{Theorem}
\newtheorem{corollary}{Corollary}[theorem]

\theoremstyle{definition}

\theoremstyle{proposition}
\newtheorem{proposition}{Proposition}[section]
\theoremstyle{assumption}
\newtheorem{assumption}{Assumption}[section]

\theoremstyle{remark}
\newtheorem{remark}{Remark}

\begin{document}


\title{Robust Path-following for Keplerian Orbits}  

\author[ a]{Rodolfo Batista Negri\footnote{email: rodolfo.negri@unifesp.br. \\ Personal website: \url{rodolfobnegri.com}}}
\author[ b]{Antônio F. B. de A. Prado}
\affil[ a]{Federal University of São Paulo, ICT, 12247-014, São José dos Campos, São Paulo, Brazil}
\affil[ b]{National Institute for Space Research, 1758, 12227-010, São José dos Campos, SP, Brazil}
\date{}


      
\maketitle

\begin{abstract}                          
This work introduces a novel path-following control strategy inspired by the famous two-body problem, aiming to stabilize any Keplerian orbit. Utilizing insights from the mathematical structure of the two-body problem, we derive a robust path-following law adopting sliding mode control theory to achieve asymptotic convergence to bounded disturbances. The resulting control law is demonstrated to be asymptotically stable. Illustrative examples showcase its applicability, including orbiting an accelerated moving point, patching Keplerian trajectories for complex patterns, and orbital maintenance around the asteroid Itokawa. The proposed control law offers a significant advantage for the orbital station-keeping problem, as its sliding surface is formulated based on variables commonly used to define orbital dynamics. This inherent alignment facilitates easy application to orbital station-keeping scenarios.
\end{abstract}

\section{Introduction}\label{sec1}

For certain operations of a vehicle, using path-following control laws is more interesting than reference tracking. Path-following algorithms focus on driving the vehicle to a specific path geometry and maintaining it without any time parameterization. In contrast, reference tracking forces the vehicle to reach a particular point on the path at a specific time. In mathematical terms, in a path-following approach, the desired output of the system is parameterized by a virtual arc-length $\theta \in [0;\theta_f]$, where $\theta_f$ is the total virtual arc-length, denoted as $\vec{y}_d(\theta)$ \cite{aguiar2005path,rubi2020survey}, while reference tracking uses $\vec{y}_d(t)$, which is parameterized on time $t$.

Path-following is useful in scenarios where the path itself is more critical than reaching certain points at precise times. This is common in scenarios like lane-keeping on highways~\cite{marino2011nested}, where the vehicle needs to accurately follow the curvature of the road. On the other hand, reference tracking is suitable for tasks that require precise timing and synchronization, such as the platooning of vehicles~\cite{stankovic2000decentralized}. Path-following control also offers significant performance advantages compared to reference tracking~\cite{aguiar2004path,aguiar2005path,aguiar2008performance}.  Aguiar et al.~\cite{aguiar2008performance} conducted a study illustrating that tracking a geometric path $y_d(\theta)$ is less restrictive than tracking a reference signal $\vec{y}_d(t)$. The latter is subject to limitations imposed by the unstable zero dynamics, limiting the ability to reduce the $\mathcal{L}_2$-norm of the tracking error to an arbitrarily small value.

Those advantages made path-following a chosen strategy for  the guidance of various types of vehicles, including aircraft \cite{consolini2010path,kai2019unified,wang2019cooperative}, watercraft \cite{belleter2019observer,zuo2018three,zheng2020moving}, and robots \cite{wit1993nonlinear,kapitanyuk2017guiding,kelasidi2017integral,wu20193}. Researchers have proposed various approaches to achieve path-following, such as using transverse feedback linearization \cite{banaszuk1995feedback,nielsen2008local,nielsen2010path}, vector field \cite{nelson2007vector,yao2020path}, and line-of-sight \cite{fossen2003line,borhaug2010straight}. A survey of different planar path-following algorithms used for guiding UAVs (unmanned aerial vehicles) can be found in \cite{sujit2014unmanned}.

Those previously proposed path-following approaches rely on intermediate steps such as path parameterization and putting the problem in terms of heading/yaw rate/angle to control the vehicle's path. In this paper, we put on a solid mathematical foundation a new approach to the path-following control, which has already demonstrated exceptional performance for operations near small celestial bodies~\cite{negri2021orbit,batista2022autonomous,negri2022autonomous}. 

{

Instead of framing the path following problem in terms of heading, yaw rate, or angle, our approach draws inspiration from the two-body problem and directly incorporates constants that define the conic section into the control law. This innovative synthesis allows for the seamless application of the method to astronautics problems, eliminating the need for intermediate steps. Furthermore, it facilitates convenient adjustments of parameters to maintain the desired orbit. To achieve this, we leverage the integrals of motion that define the orbital geometry in the two-body problem \cite{goldstein2002classical}. By controlling some of these constants, the path-following problem is formulated as a regulation problem of the angular momentum and eccentricity vectors of the two-body problem. This unique approach allows any particle to describe a Keplerian orbit in a path-following guidance manner.
}

Given the prevalence of disturbances encountered in most applications of this control law, such as solar radiation pressure and gravity field non-uniformity, we have designed the path-following law based on sliding mode control theory to achieve robust control. Initially, we represent the particle's equations of motion in a radial-transverse-normal (RTN) frame, akin to the Frenet-Serret frame. We introduce a novel approach for constructing a sliding surface, utilizing a linear combination of radial and transverse components for each vector. Proof of the asymptotic stability of this new sliding surface is provided. One key advantage of our approach is its ability to use a single sliding surface and control command for effectively controlling the plane of motion. This simplicity and efficiency make it a compelling candidate for further research in applications where only the plane of motion should be controlled.

To validate the effectiveness of our proposed control law, we showcase its performance in three distinct applications. The first application addresses the challenging moving path-following (MPF) problem \cite{oliveira2013moving,oliveira2016moving}, where we task a particle to follow an orbit around a point that is following an accelerated sinusoidal trajectory. The point's movement is considered a disturbance to the system. The second example involves a scenario with patched hyperboles, requiring the particle to approach three different checkpoints. Lastly, we demonstrate our control strategy's efficacy in the autonomous orbit-keeping problem around the asteroid Itokawa, accounting for perturbation effects like solar radiation pressure, higher-order gravity field terms, and unknown spin state. Through these applications, we present evidence of the stability and versatility of our proposed control approach for various scenarios.

\section{Problem Statement}
\label{sec:dyn}

To ensure the broad applicability of the proposed control law, we focus solely on the outer loop guidance of the vehicle's center of mass, without committing to any specific problem. For those interested in real-world applications where the proposed control law excels, we recommend exploring the results from applying the control approach of this paper to the problem of orbital maintenance around small bodies~\cite{batista2022autonomous}. 

We assume that the particle follows the following equations of motion in a reference frame centered on the point it will orbit:

\begin{subequations}
\label{eq:System}
\begin{align}
\dot{\vec{r}}(t) &= \vec{v}(t), \\
\dot{\vec{v}}(t) &= \vec{f}(\vec{r},\vec{v},t) + \vec{d}(\vec{r},\vec{v},t) + \vec{u}(\vec{r},\vec{v},t),
\end{align}
\end{subequations}
where $\vec{r}(t)$ and $\vec{v}(t) \in \mathbb{R}^{3}$ are the position and velocity vectors, respectively. $\vec{f}(\vec{r},\vec{v},t)$ is a smooth nonlinear function of $\vec{r}$ and $\vec{v}$ representing known dynamics, and $\vec{d}(\vec{r},\vec{v},t)$ is unknown or unmodelled disturbances that satisfy the condition: $\left|d_j\right| \leq D_j$, where $D_j > 0$, $j=1,2,3$, in which $d_j$ are the components of $\vec{d}$.

We will derive our control laws in the radial-transverse-normal frame (RTN), similar to the Frenet-Serret~\cite{bishop1975there}. A transposition between RTN and Frenet-Serret can be obtained by a simple rotation around their normal axis. The RTN is a right-handed coordinate system in which the radial, transverse, and normal components are defined based on the particle's kinematics. The radial component is in the direction of the position vector $\vec{r}$, the normal component is perpendicular to the osculating plane in the direction of the specific angular momentum $\vec{h}$, and the transverse component completes the right-handed frame. All of them defined in an inertial frame. Their unit vectors are expressed as functions of the particle's kinematics:

\begin{subequations}
\label{eq:RTN_unit vectors}
\begin{align}
\hat{r} &= \frac{\vec{r}}{r}, \\
\hat{\theta} &= \hat{h} \times \hat{r}, \\
\hat{h} &= \frac{\vec{h}}{h}.
\end{align}
\end{subequations}

An arbitrary vector $\vec{A} \in \mathbb{R}^{3}$, in the same arbitrary reference frame as Eqs. \ref{eq:System}, can be represented in the RTN frame as:

\begin{equation}
\label{eqn:write_in_RTN}
\vec{A}_{RTN} = \begin{bmatrix}
A_R \\
A_T \\
A_N
\end{bmatrix} = \begin{bmatrix}
\vec{A}\cdot \hat{r} \\
\vec{A} \cdot \hat{ \theta} \\
\vec{A} \cdot \hat{h}
\end{bmatrix}= \begin{bmatrix}
 \hat{r}^{\text{T}} \\
 \hat{ \theta}^{ \text{T}} \\
 \hat{h}^{ \text{T}}
\end{bmatrix} \vec{A}=
[RTN] \vec{A},
\end{equation}
where the superscript $ \text{T}$ represents the transpose, and the subscripts $R$, $T$, and $N$ stand for the radial, transverse, and normal coordinates, respectively. The term $[RTN]$ is the transformation matrix that takes from the reference frame in which $\vec{A}$ is defined to the RTN.

Using Eq. \ref{eqn:write_in_RTN}, we can write the accelerations in Eqs. \ref{eq:System} in RTN as:

\begin{equation}
\vec{a}_{RTN} = \begin{bmatrix}
f_R + u_R + d_R \\
f_T + u_T + d_T \\
f_N + u_N + d_N
\end{bmatrix}.
\end{equation}

The radial and transverse accelerations will act in the following equations of motion on the osculating plane:
\begin{subequations}
\begin{align}
\ddot{r} - r\dot{\theta}^2 &=  a_R, \\ 
2 \dot{r} \dot{\theta} + r \ddot{\theta} &= a_T,
\end{align}
\end{subequations}
in which $\theta$ is a virtual arc-length representing the particle's position, for $\theta \in [0;\theta_f]$, where $\theta_f$ is the total virtual arc-length. Noting that the specific angular momentum can be represented as $h=r^2\dot{\theta}$, we can reduce the equations of motion to:
\begin{subequations}
\begin{align}
\ddot{r} &=  \frac{h^2}{r^3} + a_R, \\ 
\label{eq:hdot}
\dot{h} &= r a_T
\end{align}
\end{subequations}

The missing normal component is responsible for varying the osculating orbital plane. Its effect can be checked by obtaining the temporal variation of the specific angular momentum:

\begin{equation}
\dot{\vec{h}} = \vec{r} \times \vec{a} = ra_T \hat{h} - ra_N \hat{\theta}.
\end{equation}
Because $\dot{\vec{h}}= \dot{h}\hat{h} + h\dot{\hat{h}}$, and using Eqs. \ref{eq:RTN_unit vectors} with a simple derivation similar to the one for obtaining polar coordinates, one can check that {the normal basis of the RTN frame} will follow the subsequent equations of motion:
\begin{subequations}
\label{eq:RTN_EoM}
\begin{align}
\dot{\hat{r}} &= \frac{h}{r^2} \hat{\theta}, \\
\dot{\hat{\theta}} &= \frac{ra_N}{h} \hat{h}-\frac{h}{r^2}\hat{r}, \\
\dot{\hat{h}}  &= - \frac{ra_N}{h} \hat{\theta}.
\end{align}
\end{subequations}
These are equivalent to the Frenet-Serret formulas but for RTN coordinates. 

Consider the arbitrary vector $\vec{A}$. Also, consider this same vector written in RTN coordinates and represented as $\vec{A}_{RTN}$. We represent the derivative of $\vec{A}_{RTN}$ with respect to time $t$, taken in each of its components, as $\frac{d}{dt}(\vec{A}_{RTN})$. While $\dot{\vec{A}}$, the time derivative of $\vec{A}$, is represented in RTN coordinates as $\dot{\vec{A}}_{RTN}$.

Thus, following Eq. \ref{eqn:write_in_RTN} and the definition above, we have that:

\begin{align}
\begin{split}
    \frac{d}{dt}(\vec{A}_{RTN}) = & \frac{d}{dt}([RTN]) \vec{A} + [RTN] \dot{\vec{A}} \\ = & \frac{d}{dt}([RTN]) \vec{A} + \dot{\vec{A}}_{RTN}.
    \end{split}
\end{align}

Now, considering Eqs. \ref{eq:RTN_EoM}, it follows that:

\begin{equation}
\label{eqn:Xrtn_dot}
\frac{d}{dt}(\vec{A}_{RTN}) = \begin{bmatrix}
\dot{A}_R \\
\dot{A}_T \\
\dot{A}_N
\end{bmatrix} = \begin{bmatrix}
\dot{\vec{A}} \cdot \hat{r} + \frac{h}{r^2} A_T \\
\dot{\vec{A}} \cdot \hat{\theta} + \frac{ra_N}{h} A_N - \frac{h}{r^2} A_R \\
\dot{\vec{A}} \cdot \hat{h} - \frac{ra_N}{h} A_T  
\end{bmatrix}
\end{equation}

\subsection{Two-body Problem Fundamentals}
\label{sec:2BP}

This section introduces the fundamental concepts of the two-body problem, which will serve as inspiration for the proposed control law. We recognize that not all readers may have a background in astronautics, and some researchers from other fields may also find the proposed control law valuable or draw inspiration from it. Therefore, it is important to present the basics of the two-body problem in a clear and concise manner. Also, understanding these foundational concepts is crucial for comprehending how the control law works and how it leverages the constants defining a conic section {to maintain the orbital geometry in a path-following manner}. For readers seeking more comprehensive details on the two-body problem, we recommend referring to specialized works in classical mechanics, celestial mechanics, and astrodynamics \cite{battin1999introduction,goldstein2002classical,de2004astronomia,arnold2007mathematical}.

In the context of the two-body problem, the equation of motion governing the movement of a body with respect to another is expressed as \cite{goldstein2002classical,de2004astronomia}:

\begin{equation}
\label{eq:EoM_2BP}
    \ddot{\vec{r}} = - \frac{\mu}{r^3} \vec{r},
\end{equation}
in which $\mu$ is the gravitational parameter of the system, and $\vec{r}$ is the distance of one of the bodies with respect to the other.

Taking the cross product of $\vec{r}$ with Eq. \ref{eq:EoM_2BP} on both sides we have that:

\begin{equation}
   \vec{r} \times \ddot{\vec{r}} = 0,
\end{equation}
which can be rewritten to find that:

\begin{equation}
   \vec{r} \times \ddot{\vec{r}} = \frac{d}{dt}( \vec{r} \times \dot{\vec{r}} ) = \frac{d \vec{h}}{dt} = 0.
\end{equation}
Thus, $\vec{h}$ is an integral of motion of Eq. \ref{eq:EoM_2BP}, and it is precisely the specific angular momentum vector $\vec{h}=\vec{r}\times \vec{v}$. Because $\vec{h}$ is constant, it follows that the movement described by Eq. \ref{eq:EoM_2BP} must lie on a plane.

Taking the cross product of the right-hand side of Eq. \ref{eq:EoM_2BP} with $\vec{h}$, it follows that:

\begin{align}
\begin{split}
    - \frac{\mu}{r^3} \vec{r} \times \vec{h} = & - \frac{\mu}{r^3} \vec{r} \times (\vec{r} \times \dot{\vec{r}} ) \\ = & - \frac{\mu}{r^3}  \left[ ( \vec{r} \cdot \dot{\vec{r}} ) \vec{r} - ( \vec{r} \cdot \vec{r} ) \dot{\vec{r}}   \right] \\ = & - \frac{\mu}{r^3}  ( \vec{r} \cdot \vec{v} ) \vec{r} +  \frac{\mu}{r}\vec{v}.
\end{split}
\end{align}

Noting that $r=(\vec{r} \cdot \vec{r})^{1/2}$, we have:
\begin{equation}
     - \frac{\mu}{r^3}  ( \vec{r} \cdot \vec{v} ) \vec{r} +  \frac{\mu}{r}\vec{v} =  \mu \frac{d}{dt} \left( \frac{\vec{r}}{r} \right)
\end{equation}

We also take the cross product with $\vec{h}$ on the left-hand side to obtain:
\begin{equation}
    \ddot{\vec{r}} \times \vec{h} = \frac{d}{dt} (\vec{v} \times \vec{h}).
\end{equation}

Therefore:
\begin{equation}
      \mu \frac{d}{dt} \left( \frac{\vec{r}}{r} \right) = \frac{d}{dt} (\vec{v} \times \vec{h})
\end{equation}
which integrating both sides yields:

\begin{equation}
\label{eq:evec_anteriorrrr}
      \frac{\vec{r}}{r}  + \vec{e} = \frac{1}{\mu} \vec{v} \times \vec{h},
\end{equation}
where $\vec{e}$ is a constant of integration, which can be rewritten to:

\begin{equation}
\label{eq:evec}
\vec{e} = \frac{1}{\mu} (\vec{v} \times \vec{h} - \mu \hat{r}).
\end{equation}
Equation \ref{eq:evec} also represents an integral of motion for Eq. \ref{eq:EoM_2BP} and is commonly known as the eccentricity or Runge-Laplace-Lenz vector \cite{goldstein2002classical}. One might expect that having both vectors, $\vec{h}$ and $\vec{e}$, would be sufficient to solve the problem, as they provide six integrals of motion, seemingly allowing us to solve the three second-order differential equations presented in Eq. \ref{eq:EoM_2BP}. However, it is important to note that:
\begin{equation}
\label{eq:he_perp}
    \vec{h} \cdot \vec{e} = \frac{1}{\mu} \vec{h} \cdot  (\vec{v} \times \vec{h})  -  \vec{h} \cdot \hat{r}=  \frac{1}{\mu} \vec{v} \cdot  (\vec{h} \times \vec{h}) = 0.
\end{equation}
Therefore, it turns out that the vectors $\vec{h}$ and $\vec{e}$ provide only five independent integrals of motion, leaving one more integral to be obtained for solving the two-body problem. This missing integral is derived from the Kepler equation, which goes beyond the scope of this work. However, for our purposes, it suffices to understand the geometric significance and constant nature of the vectors $\vec{h}$ and $\vec{e}$.

We can take the dot product of $\vec{r}$ with Eq. \ref{eq:evec_anteriorrrr} to find that:

\begin{equation}
    \vec{r} \cdot \vec{e} + r = \frac{1}{\mu} \vec{r} \cdot (\vec{v} \times \vec{h}) = \frac{1}{\mu} \vec{h} \cdot (\vec{r} \times \vec{v}) = \frac{h^2}{\mu}
\end{equation}
Let us define an arc-length $\theta$ such that $\vec{r} \cdot \vec{e} = r e \cos \theta$. Thus we can write:

\begin{equation}
r = \frac{h^2/\mu}{1+e \cos \theta},
\end{equation}
which is the equation of a conic section in polar coordinates with the origin in one of its foci. That is the proof of the first law of Kepler. In this conic section, the quantity $h^2/\mu$ represents the semi-latus rectum, and $e$ represents the eccentricity, which justifies referring to $\vec{e}$ as the eccentricity vector.

As we previously observed, since $\vec{h}$ is constant, the conic section must lie on a plane perpendicular to $\vec{h}$. Hence, we can express the vector $\vec{r}$ as a linear combination of the linearly independent unit vectors $\vec{e}/e$ and $(\vec{h} \times \vec{e})/(he)$ as follows:

\begin{align}
\label{eq:2BP_vecr}
\begin{split}
    \vec{r} = & r \left[ \cos \theta \frac{1}{e} \vec{e} + \sin \theta \frac{1}{he} (\vec{h} \times \vec{e}) \right] \\ = & \frac{h^2}{\mu e} \left[ \frac{\cos \theta}{1+e \cos \theta}  \vec{e} + \frac{1}{h} \frac{\sin \theta}{1+e \cos \theta} (\vec{h} \times \vec{e}) \right] 
\end{split}
\end{align}
which can be derived in time, knowing that $\dot{\theta} = h/r^2$, to find that:

\begin{align}
\begin{split}
    \vec{v} & =  \frac{h^2}{\mu e} \left[ - \frac{\sin \theta}{(1+e \cos \theta)^2}  \vec{e} + \frac{1}{h} \frac{\cos \theta + e}{(1+e \cos \theta)^2} (\vec{h} \times \vec{e}) \right] \dot{\theta} 
    \\ &=\frac{h^2}{\mu e} \left[ - \frac{\sin \theta}{(1+e \cos \theta)^2}  \vec{e} + \frac{1}{h} \frac{\cos \theta + e}{(1+e \cos \theta)^2} (\vec{h} \times \vec{e}) \right] \\ & \frac{(1+e\cos\theta)^2\mu^2}{h^3}.
    \end{split}
\end{align}
Thus:
\begin{equation}
\label{eq:2BP_vecv}
    \vec{v} = \frac{\mu}{h e} \left[ -\sin \theta  \vec{e} + \frac{1}{h} (\cos \theta + e) (\vec{h} \times \vec{e}) \right].
\end{equation}

Therefore, Eqs. \ref{eq:2BP_vecr} and \ref{eq:2BP_vecv} provide the solution to Eq. \ref{eq:EoM_2BP}. The evolution of the arc-length $\theta$ is governed by the last independent integral of motion related to the Kepler equation.

For the purpose of this paper, it is important to note that by controlling the vectors $\vec{h}$ and $\vec{e}$, any vehicle can follow a Keplerian motion in a path-following guidance manner. Thus, considering the system in Eqs. \ref{eq:System}, we can make its output $\vec{y}(t) = \begin{bmatrix} \vec{r}(t) & \vec{v}(t) \end{bmatrix}^\text{T}$ follow the desired geometric path $\vec{y}_d(\theta) = \begin{bmatrix} \vec{r}(\theta) & \vec{v}(\theta) \end{bmatrix}^\text{T}$, where $\vec{r}(\theta)$ and $\vec{v}(\theta)$ satisfy Eqs. \ref{eq:2BP_vecr} and \ref{eq:2BP_vecv}, by regulating $\vec{h}$ and $\vec{e}$. This is equivalent to achieving a path-following guidance with the assigned velocity:

\begin{align}
\begin{split}
    v = \vert \vert \vec{v} \vert \vert = & \frac{\mu}{h e} \left[ e^2 \sin^2 \theta   + (\cos \theta + e)^2 e^2 \right]^{1/2} \\ = & \frac{\mu}{ h} \sqrt{  1 + e^2 + 2 e \cos \theta }.
    \end{split}
\end{align}



\section{Main Results}

In order to derive the control law, we need to make the following Assumption:

\begin{assumption}
\label{assump.:hr}
The particle's specific angular momentum and position vectors relative to the point to be orbited are such that $\vec{h}\neq 0$ and $\vec{r}\neq 0$.
\end{assumption}

We can now make the proposition of a new sliding surface that will be central for our control derivation.

\begin{proposition}
\label{prop:sliding_surf}
A sliding surface $s \in \mathbb{R}$, written as a linear combination of the radial and transverse components of an arbitrary vector $\vec{A} \in \mathbb{R}^{3 }$:
\begin{equation}
\label{eq:s_RTN}
s=A_T+ \lambda A_R,
\end{equation}
$\lambda > 0$, will asymptotically converge to $A_T = A_R = 0 $, if $\dot{\vec{A}}=0$.
\end{proposition}

\begin{proof} From the first element of the vector in Eq. \ref{eqn:Xrtn_dot} it follows that:
\begin{equation}
A_T = \frac{1}{\dot{\theta}} (  \dot{A}_R - \dot{\vec{A}} \cdot \hat{r} )
\end{equation}

Because $\dot{\vec{A}}=0$, the sliding surface $s$ in Eq. \ref{eq:s_RTN} can be rewritten as:

\begin{equation}
\label{eq:sPROPO}
s = \frac{1}{\dot{\theta}} \dot{A}_R + \lambda A_R = 0,
\end{equation}
which has the following solution for the radial component of $\vec{A}$:

\begin{equation}
A_R(t) = A_{R}(t_0) e^{-\lambda \left( \theta(t) - \theta(t_0) \right)}
\end{equation}

But noting that $\dot{\theta}=h/r^2$, one find that:
\begin{equation}
\theta(t) - \theta(t_0) =  \int_{t_0}^t \frac{h(\tau)}{r^2(\tau)} d\tau,
\end{equation}
that is a monotonic increasing function, since $h$ and $r$ represent the magnitude of $\vec{h}$ and $\vec{r}$, respectively, and Assumption \ref{assump.:hr} prevents them to be zero, so $h, r>0$. Therefore, $A_R$ will asymptotically converge to $A_R=0$. In this case, $A_T=0$ follows from it.  \end{proof}

\begin{remark}
It is important to clarify that the sliding surface presented in Eq. \ref{eq:s_RTN} should not be mistaken for what some authors refer to as a ``conventional'' sliding surface \cite{shtessel2014sliding}, merely because it involves a linear combination. In the conventional form, the sliding surface is given as $s=\dot{\varepsilon}(t) + \lambda \varepsilon(t)$, where $\dot{\varepsilon}$ and $\varepsilon$ are real-valued variables representing the output error of a second-order arbitrary system \cite{shtessel2014sliding}. In contrast, the Eq. \ref{eq:s_RTN} is a linear combination of components of a constant vector $\vec{A}\in \mathbb{R}^3$. It is crucial to understand that the effectiveness of the sliding surface lies in its ability to produce the desired results when the system operates over it. In our case, the condition $A_T=A_R=0$ is particularly advantageous and will be leveraged in the subsequent analysis.
\end{remark}

Proposition \ref{prop:sliding_surf} is very important since it allows to control a vector $\vec{A}$ by a single sliding surface. We will make use of it for controlling the osculating plane. The point to be orbited is already defined by writing the equations of motion, Eqs. \ref{eq:System}, in a frame centered on it, the particle's plane is defined by the unit vector $\hat{h}$. Therefore, we can make use of this fact to choose a desired specific angular momentum unit vector $\hat{h}_d$, defined in RTN using Eqs. \ref{eqn:write_in_RTN} as: 
\begin{equation}
\hat{h}_{d-RTN} = \begin{bmatrix}
h_{dR} \\
h_{dT} \\
h_{dN}
\end{bmatrix},
\end{equation}
to make $\hat{h}$ converge to it.

\begin{assumption}
\label{assump.:Bound_beta}
Let $\beta$ be an angle between the desired angular momentum unit vector $\hat{h}_d$ and the actual angular momentum unit vector $\hat{h}$, $\cos \beta = \hat{h} \cdot \hat{h}_d$. The magnitude of this angle is bounded such that $\beta< 90 ^\circ$.
\end{assumption}

Now, we can derive a control law for controlling the osculating plane, given by the following theorem.

\begin{theorem}
\label{theor:plane_control}
Assuming that the perturbation normal to the orbit, $d_N$, is bounded such that $|d_N|<D_N$. A control normal to the osculating plane:
\begin{equation}
\label{eq:u_N}
u_N=\frac{h^2}{r^3 h_{dN}}(h_{dR}-\lambda_N h_{dT})-K_N \text{sgn}(s_N) -f_N,
\end{equation}
$K_N \geq D_N$, will guarantee convergence to the sliding surface: 
\begin{equation}
\label{eq:s_N}
 s_N = \hat{h}_d \cdot (\lambda_N \hat{r} + \hat{\theta}) = h_{dT} + \lambda_N h_{dR} = 0,
\end{equation} 
and, once reached, the osculating plane will asymptotically converge to the desired plane defined by $\hat{h}_d$.
\end{theorem}

\begin{proof} Taking the derivative of Eq. \ref{eq:s_N}, it follows that:

\begin{equation}
\dot{s}_N = \dot{h}_{dT} + \lambda_N \dot{h}_{dR}
\end{equation}
Using Eq. \ref{eqn:Xrtn_dot} and the fact that $\dot{\hat{h}}_d =0$:

\begin{equation}
\dot{s}_N = \frac{r h_{dN}}{h} a_N + \frac{h}{r^2} \left( \lambda_N  h_{dT} - h_{dR} \right).
\end{equation}
Choosing the following Lyapunov candidate:

\begin{equation}
V = \frac{1}{2}s_N^2,
\end{equation}
we find that:
\begin{equation}
\dot{V} =  s_N \dot{s}_N .
\end{equation}

Because $a_N = f_N + d_N + u_N$:
\begin{equation}
\label{eq:dotSN}
\dot{V} = s_N \left[ \frac{r h_{dN}}{h} (f_N+d_N+u_N) + \frac{h}{r^2} \left( \lambda_N  h_{dT} - h_{dR} \right) \right]
\end{equation}
Substituting the control $u_N$ given by Eq. \ref{eq:u_N} we obtain:
\begin{equation}
\dot{V} = \frac{r h_{dN}}{h} s_N (d_N-K_N \text{sgn}(s_N)) 
\end{equation}

The Assumption \ref{assump.:Bound_beta} guarantees that $h_{dN}>0$. Therefore, because $K_N \geq D_N$, and remembering Assumption \ref{assump.:hr}, the magnitude of $\dot{V}$ is bounded such that $\dot{V}<0$ for all $s_N\neq 0$. In this way, we show using the Lyapunov's second method that the system is stable in $s_N=0$. This implies, using Proposition \ref{prop:sliding_surf}, that the osculating plane will asymptotically converge to the desired plane defined by $\hat{h}_d$, since $\dot{\hat{h}}_d=0$. \end{proof}


\begin{corollary}
Finite time convergence to the sliding surface $s_N$, given by Eq. \ref{eq:s_N}, can be obtained by choosing:
\begin{equation}
\label{eq:u_N_COROL}
u_N=\frac{h^2}{r^3 h_{dN}}\left(h_{dR}-\lambda_N h_{dT} -K_N \text{sgn}(s_N) -f_N \right).
\end{equation}
\end{corollary}


\begin{remark}
Note that Assumption \ref{assump.:Bound_beta} also prevents $u_N$ from reaching the singularity $h_{dN}=0$.
\end{remark}

\begin{remark}
In practice, Assumption \ref{assump.:Bound_beta} can be circumvent easily in an algorithm that conventionally choose an intermediary $\hat{h}_d$ if $\beta \geq 90 ^\circ$.
\end{remark}

\begin{remark}
Note that Theorem \ref{theor:plane_control} decouples the osculating plane by a single sliding surface and control command. That can be useful for other works as, in theory, it allows for any planar path-following algorithm \cite{sujit2014unmanned} to be applied after the convergence of the osculating plane given by Theorem \ref{theor:plane_control}.
\end{remark}

The Proposition \ref{prop:sliding_surf} is an innovative way of approaching this problem to obtain a robust control law such as the one in Theorem \ref{theor:plane_control}.  Let us now consider the angle $\beta$ defined in Assumption \ref{assump.:Bound_beta}. One might initially think that a sliding surface $s_N=\tilde{\beta}=\beta - \beta_d$ would be more intuitive and effective for obtaining a robust control law.

By deriving $\cos \beta = \hat{h} \cdot \hat{h}d$ and using Eq. \ref{eq:RTN_EoM}, we find:
\begin{equation}
\sin\beta \dot{\beta} = 2 \frac{ra_N}{h} h_{dT}.
\end{equation}
Interestingly, this equation has no dependence on $h_{dR}$, which means that controlling the component $h_{dR}$ is impossible. This limitation is not merely due to the way the equation is derived or the chosen coordinate system; it is rooted in the definition of $\vec{h}$ itself, which is involved in the sliding surface definition. To illustrate this further, let's write $\cos \beta = \frac{1}{h} \vec{h} \cdot \hat{h}_d$ and derive it:
\begin{align}
-\sin \beta \dot{\beta} =  \frac{1}{h} \left[ (\vec{r} \times \vec{a} ) \cdot \hat{h}_d +  (\vec{r} \times \vec{v} ) \cdot \dot{\hat{h}}_d \right.
\left. - \frac{\dot{h}}{h} (\vec{r} \times \vec{v} ) \cdot \hat{h}_d  \right],
\end{align}
as we can observe, each term in the equation is dependent on the cross-product with the position vector. The sliding surface in Eq. \ref{eq:s_N}, on the other hand, effectively avoids this drawback by removing $\vec{h}$ from its definition. This crucial advantage can only be achieved using the sliding surface defined in Proposition \ref{prop:sliding_surf}.

Up to this point, we have successfully obtained a control law for the plane motion using Theorem \ref{theor:plane_control}. However, to fully control the orbit's entire geometry, further control laws are needed. As explained in Section \ref{sec:2BP}, if we aim to control a Keplerian orbit using path-following guidance, we must also regulate the magnitude of the specific angular momentum $h$ and the eccentricity vector $\vec{e}$. These additional control laws will complete the comprehensive control strategy for achieving precise path-following guidance for the entire orbit.

It can be shown that the eccentricity vector in Eq. \ref{eq:evec} can be written in RTN coordinates as:
\begin{equation}
\label{eq:eRTN}
\vec{e}_{RTN} = \frac{1}{\mu} \begin{bmatrix}
\frac{h^2}{ r} -\mu \\ 
-\dot{r} h \\
0
\end{bmatrix},
\end{equation}
and its time derivative, also written in RTN:
\begin{equation}
\label{eq:dev_eRTN}
\dot{\vec{e}}_{RTN} = \frac{1}{\mu} \begin{bmatrix}
2h a_T \\
-h a_R - \dot{r}r a_T  - \frac{\mu h}{r^2}\\
-\dot{r} r a_N
\end{bmatrix}.
\end{equation}

With these, we can choose a desired specific angular momentum magnitude $h_d$ and a desired eccentricity vector:
\begin{equation}
\vec{e}_d = \begin{bmatrix}
e_{dR} \\
e_{dT} \\
e_{dN} 
\end{bmatrix},
\end{equation}
to derive a robust path-following control law of a Keplerian orbit, as shown in Theorem \ref{theor:2}.

\begin{theorem}
\label{theor:2}
Consider the path-following control law:
\begin{equation}
\label{eq:control_theo2}
\vec{u}_{RTN}  = - F^{-1} ( G + K \text{sgn}(\vec{s})) - \vec{f}_{RTN},
\end{equation}
where $\vec{u}_{RTN} = \begin{bmatrix} u_R & u_T & u_N\end{bmatrix}^{\text{T}}$, $K\in \mathbb{R}^{3\times 3}$ is a diagonal positive definite matrix such that its elements are $K_{j,j} \geq \text{max}( |\alpha_j|)$, $j=1,2,3$, for $\vec{\alpha} = F \vec{d}_{RTN}$, the function $\text{sgn}(\vec{s}) \in \mathbb{R}^{3 }$ represents the sign function taken in each component of $\vec{s}$, and the matrices $F$ and $G$ are defined by:
\begin{subequations}
\begin{align}
F &= \frac{1}{h\mu} \begin{bmatrix}
-h^2 & (2\lambda_R h-\dot{r}r)h & -\mu r e_{dN} \\
0 & \mu rh & 0 \\
0 & 0& \mu rh_{dN}
\end{bmatrix}, \\
G &=\frac{h}{r^2} \begin{bmatrix}
\lambda_R\tilde{e}_T-\tilde{e}_R -1 \\
0 \\
\lambda_N h_{dT} - h_{dR}
\end{bmatrix},
\end{align}
\end{subequations}
$\tilde{e}_R=e_R-e_{dR}$, $\tilde{e}_T=e_T-e_{dT}$ , and $\lambda_N, \lambda_R>0$\footnote{Note that $\dot{r}$ is the radial velocity, which is simply: $\dot{r} = \vec{v}\cdot\hat{r}$}. The control law in Eq. \ref{eq:control_theo2} will guarantee convergence to the sliding surface:
\begin{equation}
\label{eq:s_vec}
\vec{s} = \begin{bmatrix}
\tilde{\vec{e}} \cdot (\lambda_R \hat{r} + \hat{\theta}) \\
\tilde{h} \\
\hat{h}_d \cdot (\lambda_N \hat{r} + \hat{\theta})
\end{bmatrix}=0,
\end{equation}
$\tilde{h}=h-h_d$, $\tilde{\vec{e}}=\vec{e}-\vec{e}_d$ ,and, once reached, will asymptotically converge to the desired orbit geometry defined by $\vec{h}_d$ and $\vec{e}_d$.
\end{theorem}

\begin{proof} Choosing the Lyapunov candidate:
\begin{equation}
V = \frac{1}{2}  \vec{s} \cdot \vec{s},
\end{equation}
its derivative can be readily obtained:
\begin{equation}
\dot{V} = \vec{s}\cdot \dot{\vec{s}}.
\end{equation}
Thus, using Eq. \ref{eq:RTN_EoM} and knowing that $\vec{e}_d$ and $h_d$ are constants:
\begin{equation}
\dot{V} = \vec{s} \cdot \begin{bmatrix}
\dot{\vec{e}} \cdot (\lambda_R\hat{r}+\hat{\theta}) + \tilde{\vec{e}} \cdot \left(\lambda_R \frac{h}{r^2} \hat{\theta} - \frac{h}{r^2} \hat{r} + \frac{r}{h} a_N \hat{h} \right) \\
\dot{h} \\
\frac{r h_{dN}}{h} a_N + \frac{h}{r^2} (\lambda_N h_{dT} - h_{dR})
\end{bmatrix},
\end{equation}
which with Eqs. \ref{eq:hdot}, \ref{eq:eRTN} and \ref{eq:dev_eRTN}, can be written in matrix form as:
\begin{equation}
\dot{V} = \vec{s} \cdot \left( F \vec{f}_{RTN} + F \vec{u}_{RTN} + F \vec{d}_{RTN} + G \right).
\end{equation}
One can easily check that $F$ is a non-singular matrix. Because $K$ can be chosen such that $K_{j,j} \geq \text{max}(|\alpha_j|)$, $j=1,2,3$, for $\vec{\alpha} = F \vec{d}_{RTN}$, one find by applying Eq. \ref{eq:control_theo2} that $\dot{V}<0$. Therefore, the system is asymptotically stable in $\vec{s}=0$.

On the sliding surface, the second element of the sliding surface vector, $\vec{s}$, only admits the solution $h=h_d$. The third element will converge asymptotically as already shown in Theorem \ref{theor:plane_control}. Lastly, the first element can be shown to converge to $\vec{e}_d$ by using Eqs. \ref{eqn:Xrtn_dot} and \ref{eq:dev_eRTN} to rewrite it as:
\begin{equation}
\label{eq:tilde_eR}
\dot{\tilde{e}}_R= \frac{2h}{\mu}a_T - \lambda_R \dot{\theta} \tilde{e}_R.
\end{equation}
Using Filippov's Method \cite{slotine1991applied,utkin2017sliding} we can make $\dot{\vec{s}}=0$ to find that the equivalent $\vec{a}_{RTN}$ is:
\begin{equation}
\label{eq:aRTN-eq}
\vec{a}_{RTN-eq} = - F^{-1} G .
\end{equation}
By solving Eq. \ref{eq:aRTN-eq}, we can deduce that the transverse component $a_{T-eq}$ equals zero. Consequently, according to Eq. \ref{eq:sPROPO} in Proposition \ref{prop:sliding_surf}, the radial and transverse components of $\tilde{\vec{e}}$ will asymptotically converge to zero. Simultaneously, the normal component $\tilde{e}_N$ will also converge to zero, in line with the asymptotic convergence of the orbital plane. This occurs because when the orbital plane converges ($\hat{h}=\hat{h}_d$), the vectors $\vec{e}$ and $\vec{e}_d$ will solely have radial and transverse components, as they are perpendicular to $\hat{h}$ and $\hat{h}_d$ (Eq. \ref{eq:he_perp}).  \end{proof}

\begin{corollary}
\label{cor:K_matrix}
If the components of the disturbance $\vec{d}_{RTN}$ are bounded such that $\lvert d_R \rvert < D_R$, $\lvert d_T \rvert < D_T$, and $\lvert d_N \rvert < D_N$, each element of the diagonal matrix $K$ can be chosen, while still guaranteeing stability of Theorem \ref{theor:2}, accordingly to:

\begin{subequations}
\begin{align}
K_{1,1} &\geq  \frac{h}{\mu} D_R + \left\lvert \frac{2\lambda_Rh-\dot{r}r}{\mu} \right\rvert D_T + \frac{ r \left\lvert e_{dN}\right\rvert}{h}  D_N, \\
K_{2,2} &\geq r D_T, \\
K_{3,3} &\geq r \frac{h_{dN}}{h} D_N.
\end{align}
\end{subequations}
\end{corollary}

\begin{proof} Following the definition of $\vec{\alpha}$ in Theorem \ref{theor:2}, $\vec{\alpha} = F \vec{d}_{RTN}$, we obtain:
\begin{equation}
\vec{\alpha} = \begin{bmatrix}
- \frac{h}{\mu}d_R+ \frac{2\lambda_R h-\dot{r}r}{\mu}d_T-\frac{1}{h}r e_{dN} d_N\\
r d_T \\
r\frac{h_{dN}}{h} d_N
\end{bmatrix},
\end{equation}
because $K_{j,j} \geq \text{max} (\lvert \alpha_j \rvert) $, it easily follows that $K_{2,2} \geq r D_T$ and $K_{3,3} \geq r \frac{h_{dN}}{h} D_N$. For the radial component $\alpha_R$, the following bounds are true:
\begin{subequations}
\begin{align}
\alpha_R &\leq  \frac{h}{\mu} D_R + \left\lvert \frac{2\lambda_Rh-\dot{r}r}{\mu} \right\rvert D_T + \frac{ r \left\lvert e_{dN}\right\rvert}{h}  D_N = \alpha_R^+, \\
\alpha_R &\geq  - \frac{h}{\mu} D_R - \left\lvert \frac{2\lambda_Rh-\dot{r}r}{\mu} \right\rvert D_T - \frac{ r \left\lvert e_{dN}\right\rvert}{h}  D_N = \alpha_R^-.
\end{align}
\end{subequations} 
Therefore, since $\alpha_R^+ = - \alpha_R^-$, it follows that $\text{max}(\lvert \alpha_R \rvert) \leq  \alpha_R^+$, and $K_{1,1}$ can then be safely chosen respecting the bound: $K_{1,1} \geq \alpha_R^+$.  \end{proof}

It is widely known that the main drawback of the sliding mode is its discontinuous control input, which in many practical applications leads to chattering. That can be quickly dealt with by allowing the system to converge to a boundary around the sliding surface, at the expanse of a bit of performance, as extensively documented in the literature \cite{slotine1991applied,utkin2017sliding}. The most common approach is to substitute the sign function with the saturation function:
\begin{equation}
\label{eq:sat}
\text{sat}(s_j,\Phi_j) = \begin{cases} 1, &\text{$s_j>\Phi_j$} \\\frac{s_j}{\Phi_j}, &\text{$-\Phi_j \leq s_j\leq \Phi_j$} \\ -1, &\text{$s_j<-\Phi_j$} \end{cases}.
\end{equation}
In this way, we can replace the $\text{sgn}(\vec{s})$ in Eq. \ref{eq:control_theo2} by $\text{sat}(\vec{s},\vec{\Phi})$, representing the saturation function of Eq. \ref{eq:sat} taken in each component of $\vec{s}$ for the corresponding component of $\vec{\Phi}$.

{
In certain applications, allowing the orbit to vary within specified limits before engaging the control command proves beneficial~\cite{batista2022autonomous}~\footnote{{This approach is particularly advantageous for the orbital station-keeping problem in astronautics, as it accommodates periodic perturbations without requiring immediate control interventions, focusing instead on correcting secular perturbations. The formulation of the path-following law in terms of the constants of motion of the two-body problem further enhances its attractiveness, enabling simpler analysis.}}. In such scenarios, it becomes essential to ensure a minimum acceptable proximity between the spacecraft and the target orbital point. To address this need, we introduce a proposition that guarantees the maintenance of the position magnitude $r$ within predetermined bounds.
}

\begin{proposition}
If the sliding surface in Eq. \ref{eq:s_vec} is allowed to vary respecting the bounds:
\begin{equation}
\label{eq:bounds_s}
\vec{s}^- \leq \vec{s} \leq \vec{s}^+,
\end{equation}
where the $\vec{s}^-$ and $\vec{s}^+$ have all of their components negative and positive, respectively. It is guaranteed that:
\begin{equation}
\frac{(s_2^- + h_d)^2}{\mu(\lambda_R s_1^+ + e_d + 1)} \leq r \leq \frac{(s_2^+ + h_d)^2}{\mu(\lambda_R s_1^- +  1)},
\end{equation}
where the subscripts 1 and 2 represents the first and second components of the vectors $\vec{s}^-$ or $\vec{s}^-$, and $e_d = \vert \vert \vec{e}_d \vert \vert $.
\end{proposition}

\begin{proof} If the Eq. \ref{eq:control_theo2} is applied when $\vec{s}^- \leq \vec{s} \leq \vec{s}^+$, the sliding surface $\vec{s}$ is guaranteed to continue within the bounds, because we have:
\begin{equation}
\dot{\vec{s}} = -F^{-1} K \text{sgn}(\vec{s}) + F \vec{d}_{RTN} = - \Gamma(t) \text{sgn}(\vec{s}),
\end{equation}
where $\Gamma(t) \in \mathbb{R}^{3\times 3}$ is a diagonal positive definite matrix. 

One can isolate and write the radial component in Eq. \ref{eq:eRTN} as:
\begin{equation}
r = \frac{h^2}{\mu (e_R + 1)}.
\end{equation}
Applying the definition of the first and second components of the sliding surface in Eq. \ref{eq:s_vec}, $s_1$ and $s_2$ respectively, to the equation above, one find that:
\begin{equation}
r = \frac{(s_2+h_d)^2}{\mu (\lambda_R s_1 + \vec{e}_d \cdot \hat{r} + 1)}.
\end{equation}
Finally, it is easy to check that the distance $r$ respect the following conservative boundaries:
\begin{equation}
\frac{(s_2^- + h_d)^2}{\mu(\lambda_R s_1^+ + e_d + 1)} \leq r \leq \frac{(s_2^+ + h_d)^2}{\mu(\lambda_R s_1^- +  1)}.
\end{equation}
\end{proof}

\begin{remark}
In hard real-time control systems, any enhancement in computational processing time holds significant value. In this regard, it is noteworthy that the proposed control law is entirely analytical, a crucial advantage when compared to alternative approaches employing model predictive control or h-infinity methods, {which require numerical computation}. The matrix $F$ can be easily inverted analytically:\
\begin{equation}
F^{-1} = \frac{1}{r h_{dN}h} \begin{bmatrix}
-\mu r h_{dN}& h_{dN}(2\lambda_R h - \dot{r} r ) & -\mu r e_{dN} \\
0 & h_{dN} h & 0 \\
0 & 0 & h^2
\end{bmatrix} .
\end{equation}
\end{remark}

\section{Illustrative Examples}

In this section, we present simple and generic illustrative examples to demonstrate the wide applicability and versatility of the proposed path-following control law. These examples aim to cater to a diverse audience from various application fields, highlighting the control law's effectiveness in different scenarios. By avoiding specific applications, we emphasize the fundamental capabilities and robustness of the control law. {For a parametric analysis investigation and a discussing comparing this path-following law to others}, we refer readers to the application on the operation around small bodies in Negri and Prado \cite{batista2022autonomous} There, a comprehensive examination of the control law's performance with varying parameters, including $K$ and $\lambda$, is presented.

\subsection{Moving Path-following Example}

In our first example, we demonstrate the versatility and wide applicability of the proposed path-following control law through a moving path-following (MPF) problem. The objective is to showcase the effectiveness of the control law without being constrained to any specific application. In this example, a particle is tasked with orbiting a point that follows a sinusoidal trajectory with a velocity of $\vec{V}=\begin{bmatrix} 5 & \frac{5}{3} \cos \left( \frac{t}{6} \right) & 50 \cos \left( \frac{t}{7} \right) \end{bmatrix}^ \text{T}$ m/s in an inertial frame. The desired specific angular momentum and eccentricity vectors, relative to the moving point, are $\vec{h}_d=\begin{bmatrix} 0 & -8,885.8 & 8,885.8 \end{bmatrix}^{ \text{T}}$ m$^2$/s and $\vec{e}_d =\begin{bmatrix} 0 & -0.4243 & -0.4243 \end{bmatrix}^{ \text{T}}$, respectively. The control is considered to be the only known force acting on the particle, while all other forces are considered disturbances, including terms related to the accelerated point and a constant acceleration of magnitude $\begin{bmatrix} 0 & 0 & -3 \end{bmatrix}^ \text{T}$ m/s$^2$.

To evaluate the path-following law's robustness, we introduce a scenario where the particle loses all control commands for 15 seconds during the simulation. The control components are saturated by 20 m/s$^2$ in each inertial frame direction. This saturation forces the particle to recover its trajectory after the control is reinstated. The gain matrix $K$ is computed following the equality in Corollary \ref{cor:K_matrix} with $D_R=D_T=D_N=10$ m/s$^2$, and the vector $\vec{\Phi}$ is set as 5\% of each element of matrix $K$. We assign $\lambda_R=\lambda_N=2$, and $\mu$ is determined based on a desired orbital period of 100 seconds, calculated using Kepler's third law. This example can be visualized as a simplified scenario where a reconnaissance UAV is monitoring a car following a road, with the orbited point representing the car's projection at a certain height.

\begin{figure}[htb!]
\centering
\subfloat[Inertial frame]{\includegraphics[width=.8\columnwidth]{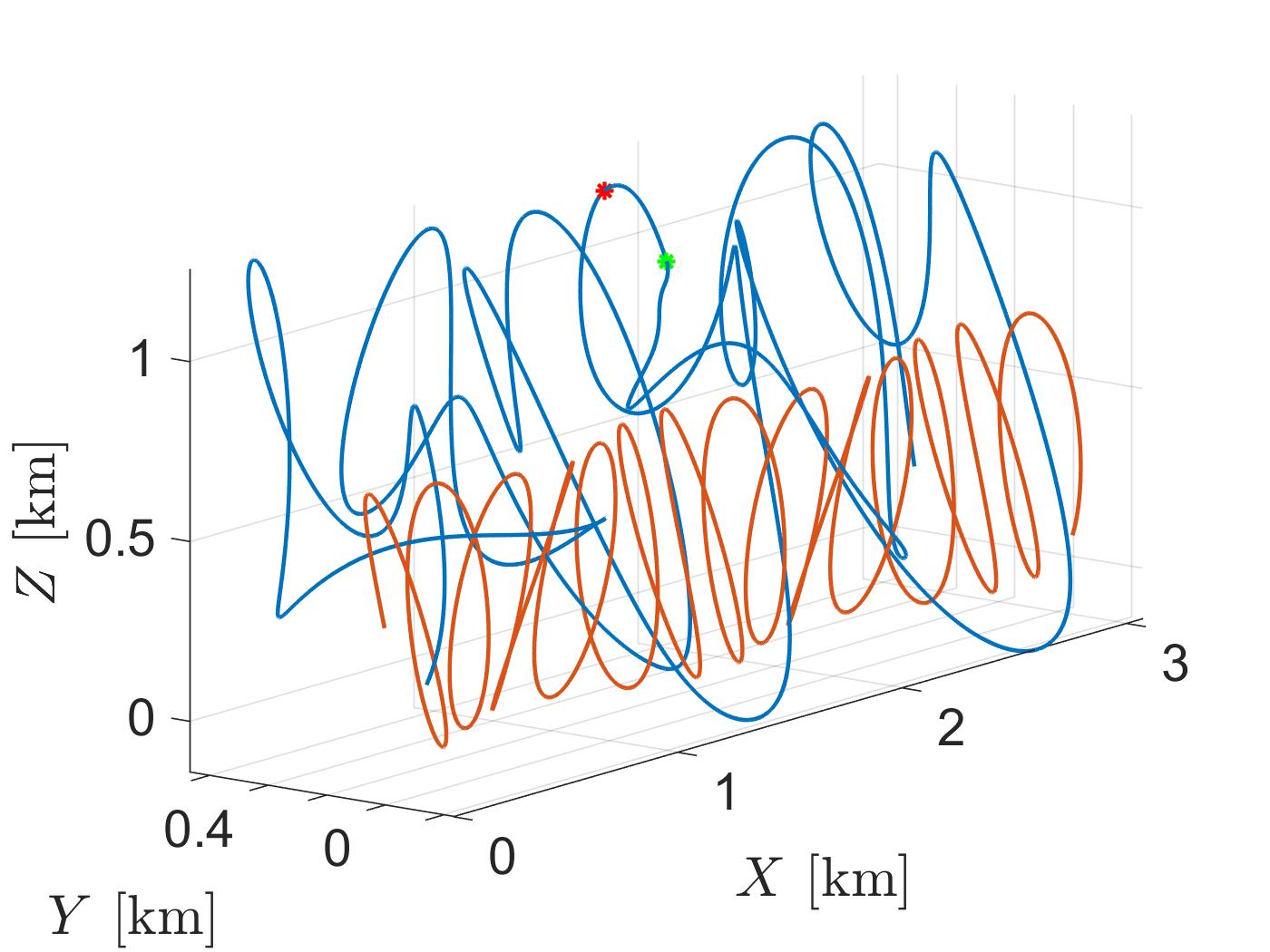}\label{Fig1a}}  \\
\subfloat[Frame centred in the orbited point]{\includegraphics[width=.8\columnwidth]{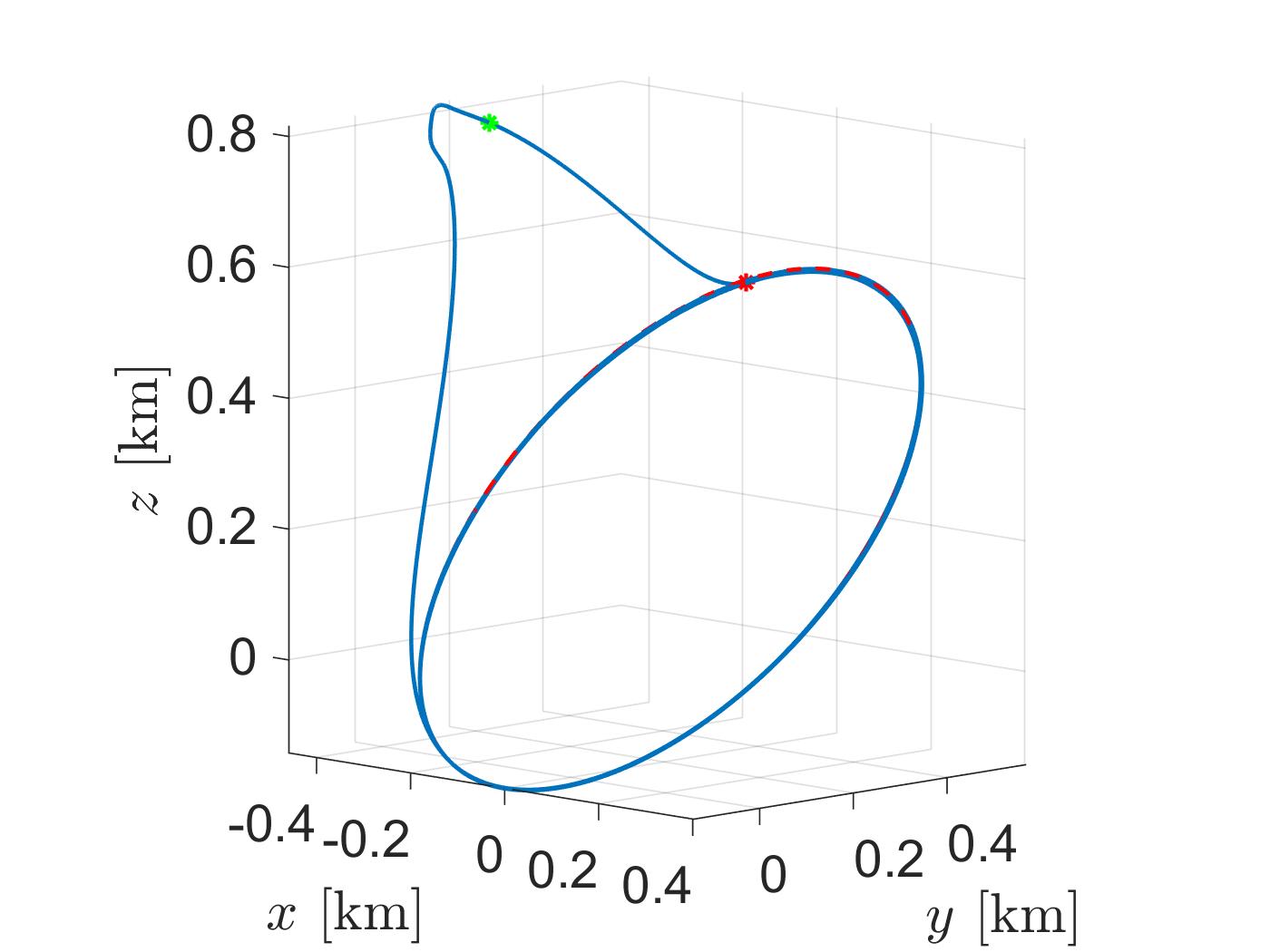}\label{Fig1b}}

\caption{Controlled orbit for the MPF example.}
\label{Fig1}
\end{figure}

\begin{figure}[htb!]
\centering
\includegraphics[width=\columnwidth]{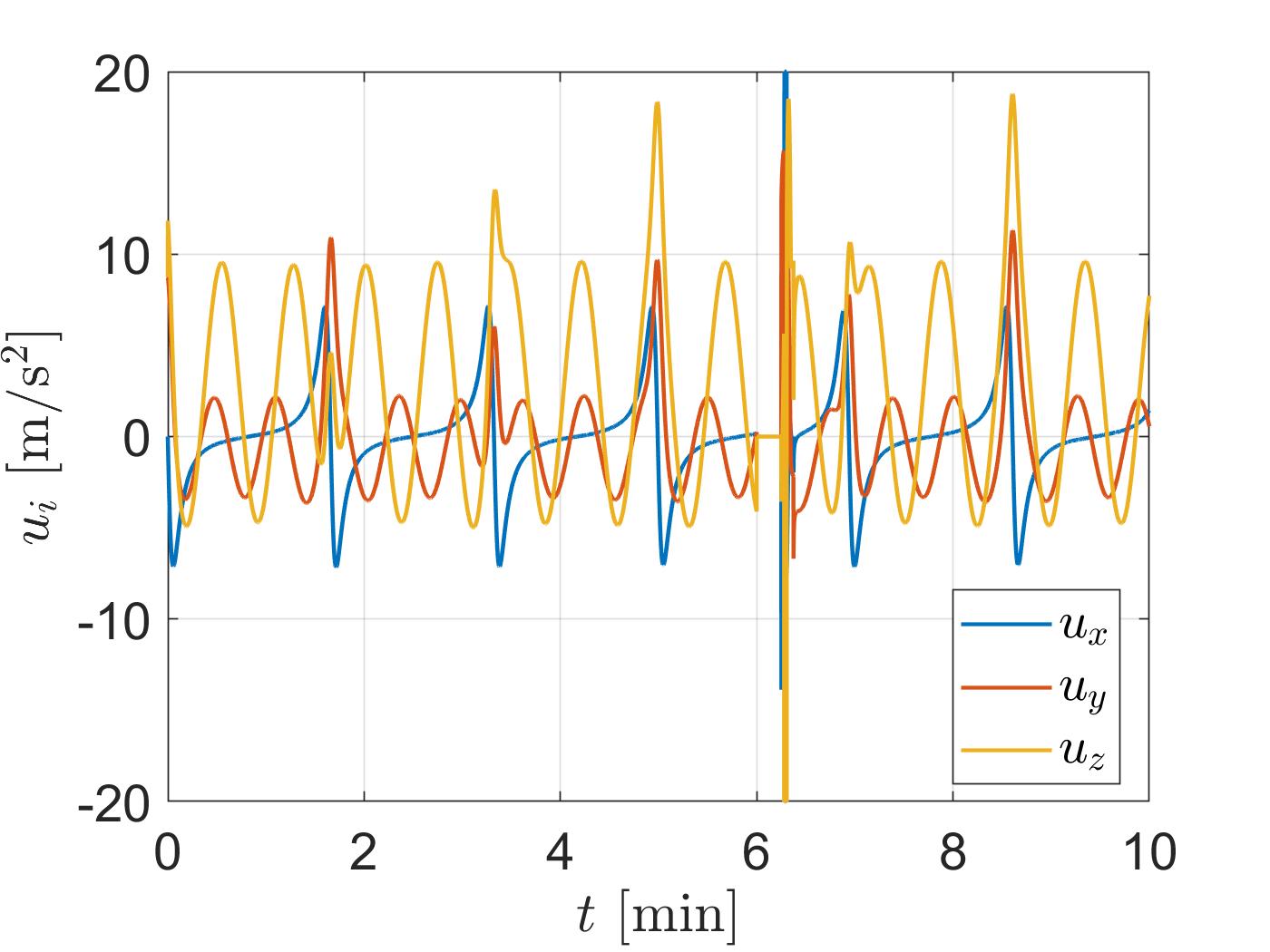}
  \caption{The control components for the MPF example.}
 \label{Fig2}
 \end{figure}

Figures \ref{Fig1} and \ref{Fig2} display the results obtained from the 10-minute simulation of the MPF example. In Fig. \ref{Fig1a}, the trajectory of the moving point is depicted in orange, representing its trajectory in the inertial frame, while the particle's trajectory is shown in blue in Figs. \ref{Fig1a} and \ref{Fig1b}. The latter figure illustrates the orbit relative to the moving point. The red asterisk marks the point in the trajectory where the particle loses control commands, and the green asterisk denotes the point where control is restored, demonstrating the control law's effective path recovery. In Fig. \ref{Fig2}, the moment at which the control command is turned off (at 6 minutes) is evident, followed by the control's prompt reaction to recover the path after it is reactivated. The effectiveness of the saturation function in Eq. \ref{eq:sat} is demonstrated in Fig. \ref{Fig2}, as it effectively removes chattering with minimal loss of performance. This is evident in Fig. \ref{Fig1b}, where the particle's orbit aligns perfectly with the desired path (indicated by a red dashed line), highlighting the high performance of the control law. We also provide a video of this simulation on the main author's webpage (\url{https://www.youtube.com/watch?v=aKbhX5QIlBs}) for the reader's reference.

\subsection{Patched-hyperboles Example}

The proposed control law demonstrates the capability to stabilize any Keplerian motion. To illustrate this, we present an example using hyperbolic trajectories. In this scenario, the particle is tasked with following a hyperbolic path relative to three different checkpoints, dynamically choosing the closest one. This type of control application could be applicable, for instance, to an UAV performing transmission line inspection.

\begin{figure}[htb!]
\centering
{\includegraphics[width=\columnwidth]{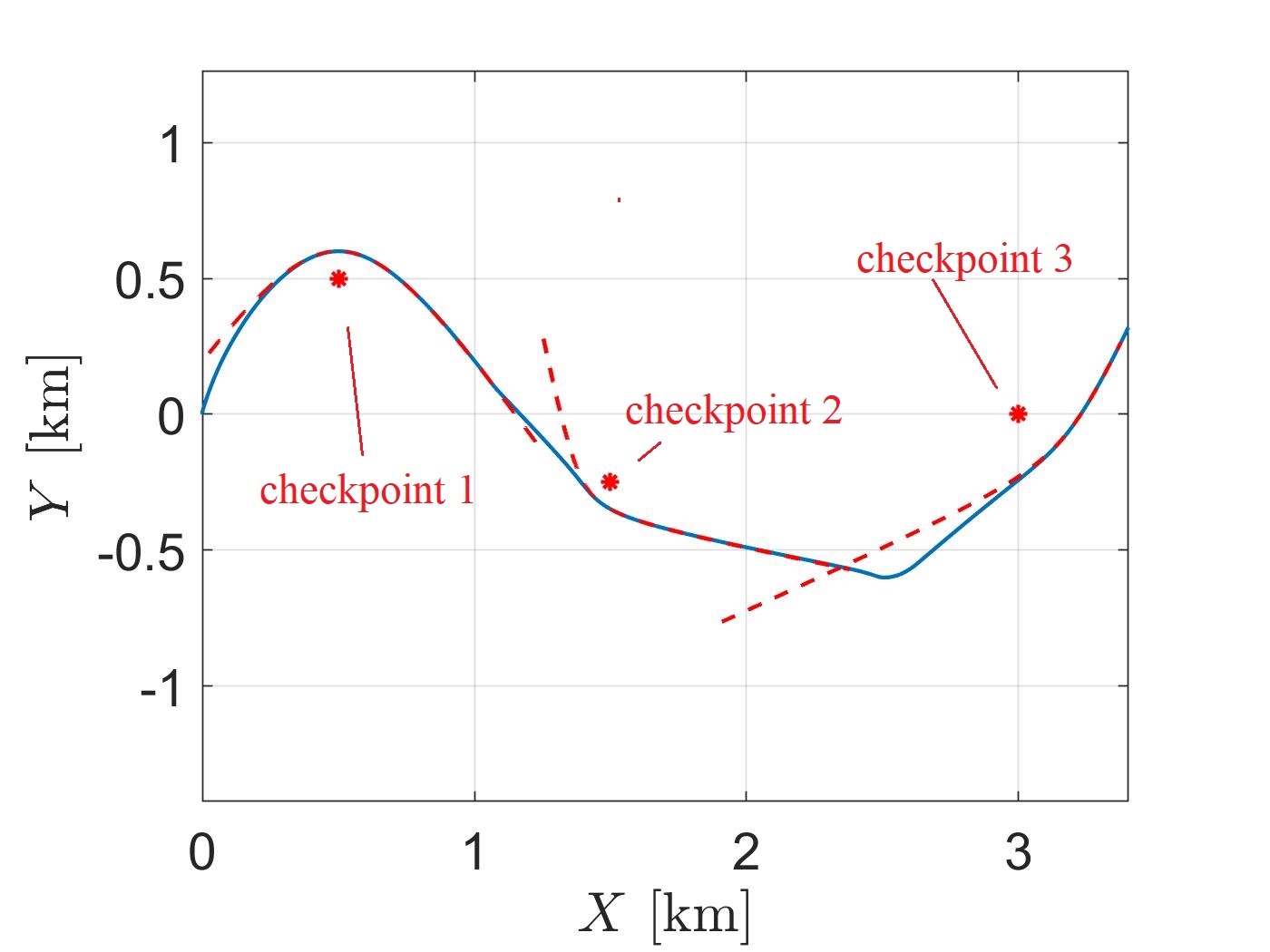}} 

\caption{Patched hyperboles example in the inertial frame.}
\label{Fig3}
\end{figure}

Figure \ref{Fig3} depicts the trajectory of the particle in an inertial frame, with each checkpoint and its respective hyperbolic path shown in red. To aid visualization, we arrange all hyperbolas on the same plane. For this example, we set $D_R=D_T=D_N=10$ m/s$^2$, and $\vec{\Phi}$ is set to 5\% of the diagonal of $K$. We also choose $\lambda_R=\lambda_N=2$ for the control parameters. As for disturbances, we consider an acceleration $\vec{d}=\begin{bmatrix} 5\sin(t) & 5\cos(t/3) & -3+5\sin(t/5) \end{bmatrix}^ \text{T}$ m/s$^2$.

\begin{figure}[htb!]
\centering
\includegraphics[width=\columnwidth]{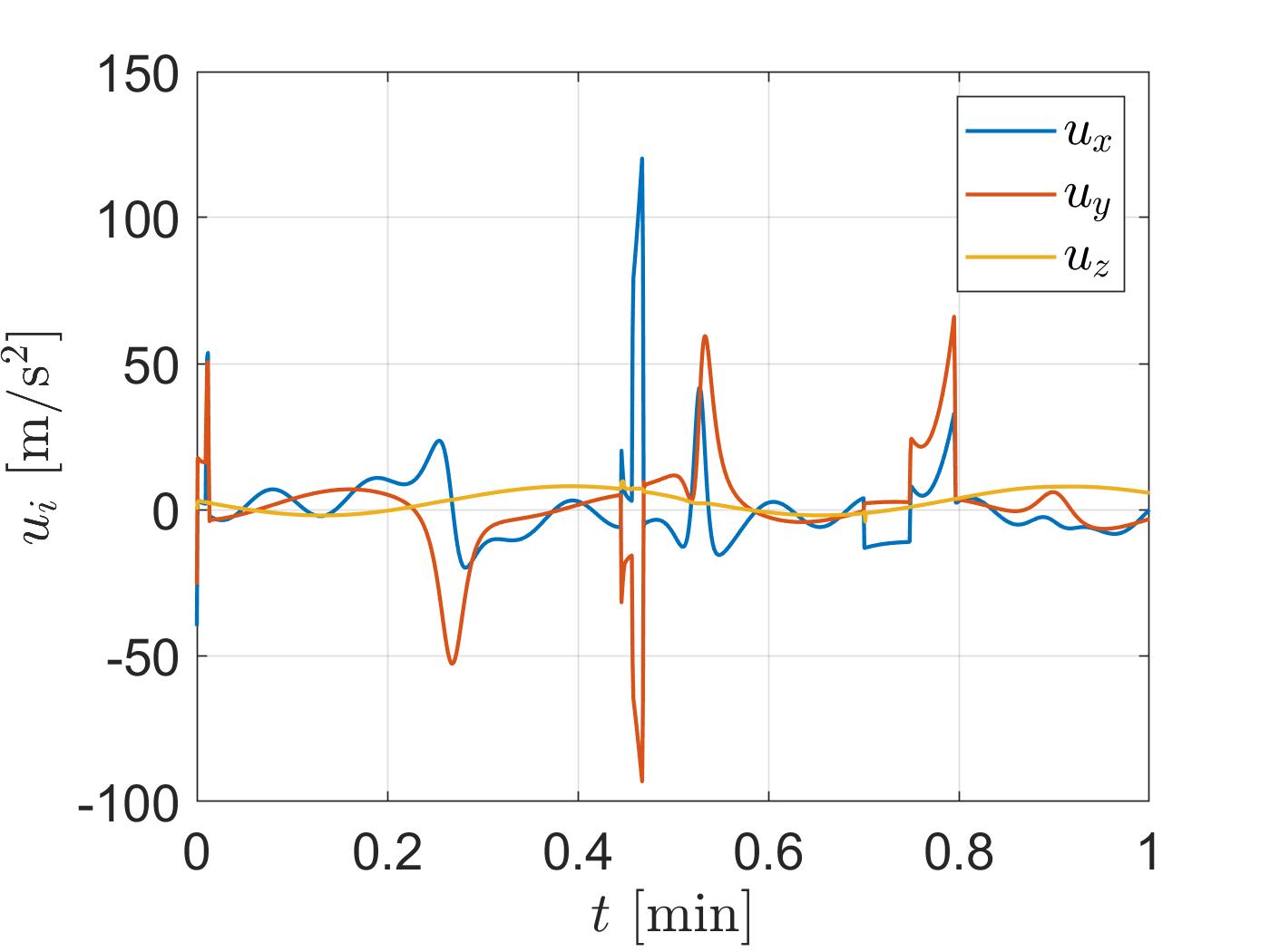}
  \caption{The control components for the patched hyperboles example.}
 \label{Fig4}
 \end{figure}

In Figure \ref{Fig4}, we present the control components used to achieve the desired hyperbolic trajectories. It is worth noting that in Figure \ref{Fig3}, the desired hyperbolic paths' legs are intentionally mismatched to stress the performance of the control law. In practical applications, the hyperbolas would be conventionally chosen to almost match their legs for a smoother transition between different checkpoints.

This example illustrates the control law's ability to handle complex trajectories, dynamically adjusting to different checkpoints, and demonstrates its potential for real-world applications in various domains where precise and adaptable path-following is essential.
 
 \subsection{Orbital Maintenance About Itokawa}

The final example focuses on the autonomous orbit-keeping problem, specifically using the asteroid Itokawa as the body to be orbited. The goal is to control an orbit corresponding to the desired values of the specific angular momentum $\vec{h}_d=\begin{bmatrix} 28.4818 & 0 & 0 \end{bmatrix}^{ \text{T}}$ m$^2$/s and the eccentricity vector $\vec{e}_d =\begin{bmatrix} 0 & 0 & 0.1 \end{bmatrix}^{ \text{T}}$. The environmental disturbances considered in this case include solar radiation pressure, the asteroid's unknown spin state, and higher-order terms of the gravity field.

\begin{figure}[htb!]
\centering
\subfloat[Uncontrolled orbit]{\includegraphics[width=.85\columnwidth]{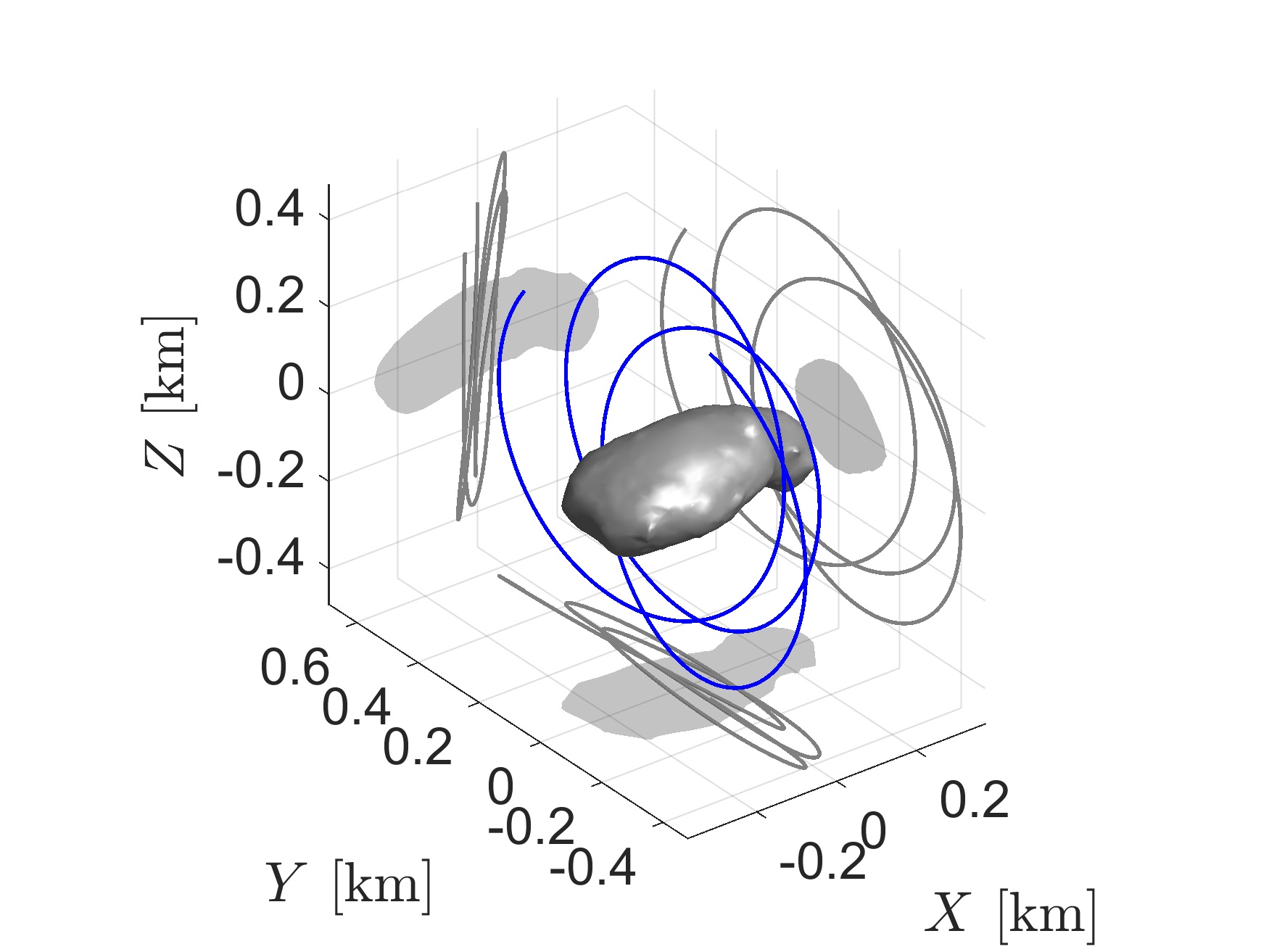}\label{Fig5a}} \\
\subfloat[Controlled orbit]{\includegraphics[width=.85\columnwidth]{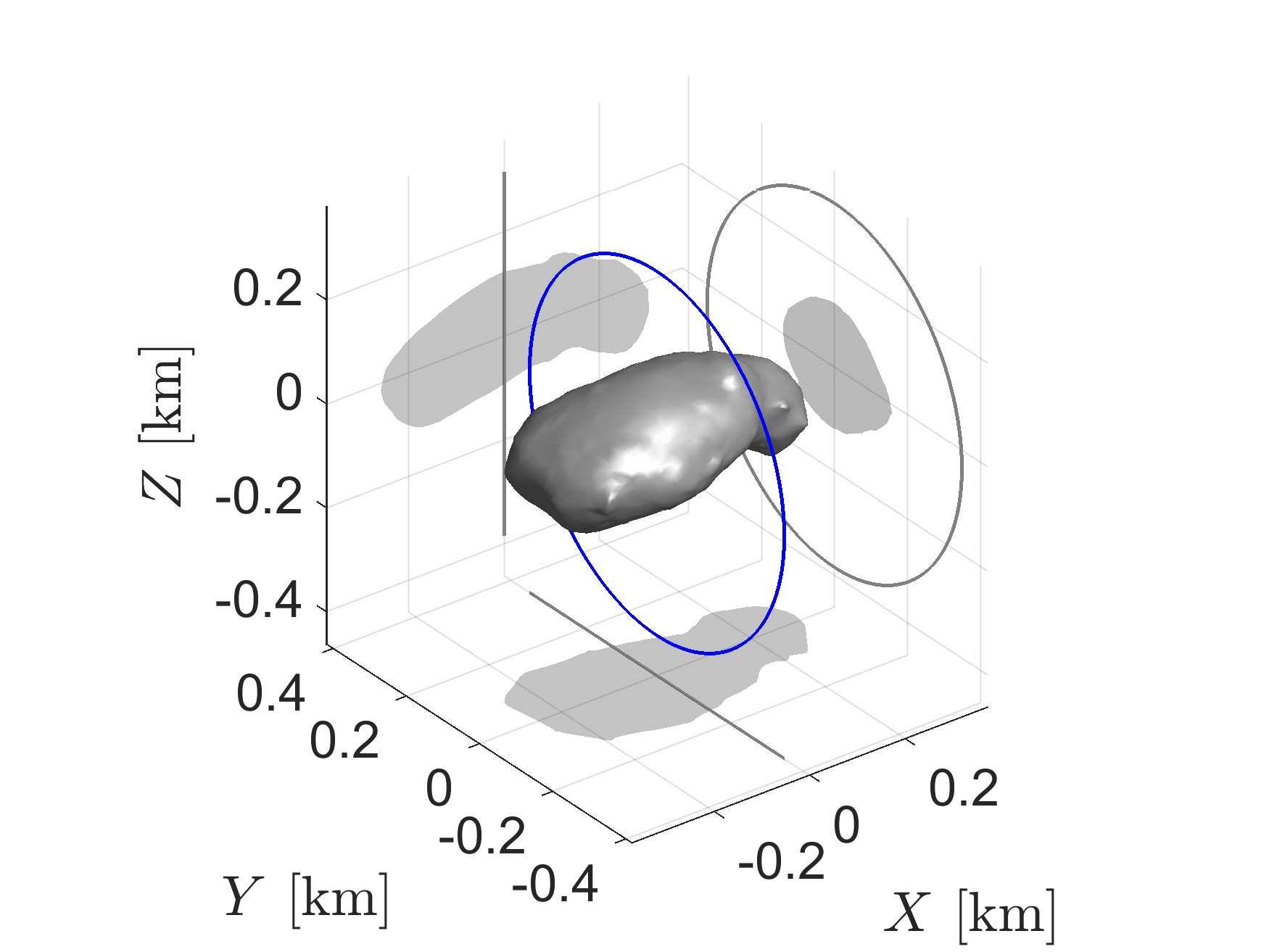}\label{Fig5b}}

\caption{Itokawa example in the inertial frame.}
\label{Fig5}
\end{figure}

\begin{figure}[htb!]
\centering
\includegraphics[width=\columnwidth]{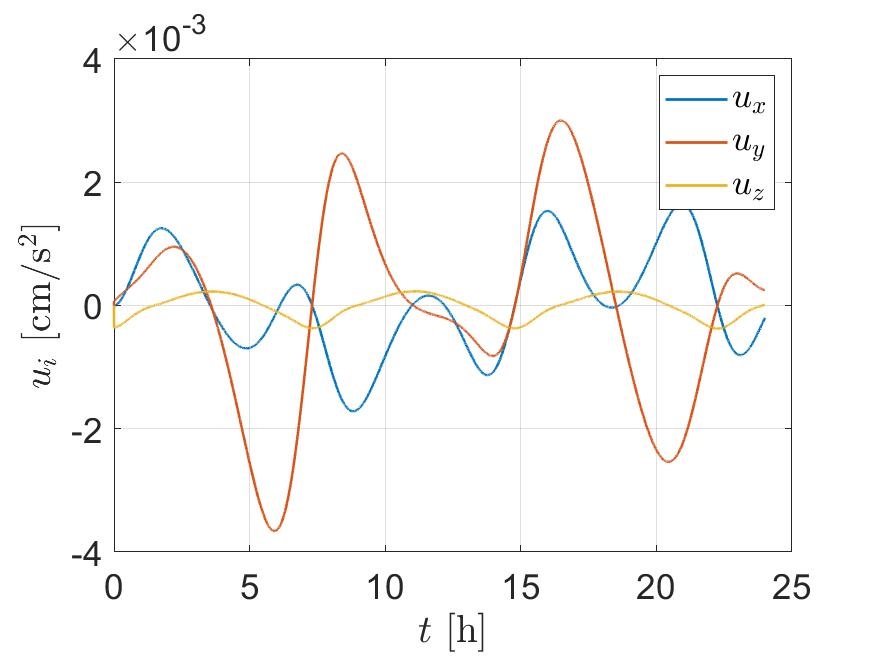}
  \caption{The control components for the Itokawa example.}
 \label{Fig6}
 \end{figure}

Figures \ref{Fig5} to \ref{Fig6} present the simulation results of a 24-hour orbit-keeping scenario in the inertial frame. The control law's performance is evident in Figure \ref{Fig5b}, showing that the orbit is successfully maintained with high accuracy. The control commands in the inertial coordinates, depicted in Figure \ref{Fig4}, demonstrate the effectiveness of the saturation function in avoiding chattering.

For a more comprehensive understanding of the modeling details, specificities, and the advantages of our path-following control law for the orbit-keeping problem, we refer the reader to Negri and Prado \cite{batista2022autonomous}. There, a detailed and focused analysis of the orbit-keeping problem is presented, considering various practical aspects that enhance the overall applicability and effectiveness of the proposed control approach. This example showcases the control law's robustness and versatility, making it suitable for real-world applications where precise and autonomous orbit-keeping is crucial, such as space missions around small celestial bodies like asteroids.

\section{Conclusions}

This work introduces a novel and robust path-following control law that can stabilize any Keplerian orbit. Leveraging the structure of the two-body problem solution, we employed sliding mode control theory to achieve asymptotic convergence. The proposed sliding surface, represented as a linear combination of radial and transverse components of a vector, enables efficient control of the vector using a single sliding surface and control command.

The decoupled osculating plane control, achieved through this approach, simplifies the robust control of the plane. Our work also serves as a foundation for future theoretical investigations, that could explore different constraints to produce non-Keplerian motion, where the proposed sliding surface could be a valuable tool. Overall, the proposed control law offers a reliable and analytical approach, making it particularly valuable in real-time control systems where computational processing time is critical. 

The fundamental attractiveness of this control law lies in its ease of application to astronautical problems, where it has already shown excellent performance. Applications include control around Martian moons \cite{negri2021orbit}, close-proximity operations around small bodies \cite{batista2022autonomous,negri2022autonomous}, and maintaining course in an autonomous gravity assist scenario \cite{negri2022study}.

Three illustrative applications were presented to demonstrate the applicability and versatility of the path-following control law. The general moving path-following problem showcased the control strategy's potential for various scenarios, such as aerial or underwater vehicles orbiting a point projected from the path of a car or ship at a specific altitude or depth. The patched hyperboles example exemplified its use in tasks like UAV transmission line inspection. Lastly, the autonomous orbit-keeping problem demonstrated its effectiveness in maintaining a spacecraft's orbit around the asteroid Itokawa, even in a highly perturbed environment.

\section*{Acknowledgements}
The authors wish to express their appreciation for the support provided by grants $\#$ 2016/24561-0, 2017/20794-2 and 2021/10853-7 from S\~ao Paulo Research Foundation (FAPESP) and the financial support from the Coordination for the Improvement of Higher Education Personnel (CAPES).  

\bibliographystyle{plain}        
\bibliography{references}        

\begin{thebibliography}{10}

\bibitem{aguiar2004path}
A~Pedro Aguiar, Dragan~B Da{\v{c}}i{\'c}, Joao~P Hespanha, and Petar
  Kokotovi{\'c}.
\newblock Path-following or reference tracking?: An answer relaxing the limits
  to performance.
\newblock {\em IFAC Proceedings Volumes}, 37(8):167--172, 2004.

\bibitem{aguiar2005path}
A~Pedro Aguiar, Joao~P Hespanha, and Petar~V Kokotovic.
\newblock Path-following for nonminimum phase systems removes performance
  limitations.
\newblock {\em IEEE Transactions on Automatic Control}, 50(2):234--239, 2005.

\bibitem{aguiar2008performance}
A~Pedro Aguiar, Jo{\~a}o~P Hespanha, and Petar~V Kokotovi{\'c}.
\newblock Performance limitations in reference tracking and path following for
  nonlinear systems.
\newblock {\em Automatica}, 44(3):598--610, 2008.

\bibitem{arnold2007mathematical}
Vladimir~I Arnold, Valery~V Kozlov, and Anatoly~I Neishtadt.
\newblock {\em Mathematical aspects of classical and celestial mechanics},
  volume~3.
\newblock Springer Science \& Business Media, 2007.

\bibitem{banaszuk1995feedback}
Andrzej Banaszuk and John Hauser.
\newblock Feedback linearization of transverse dynamics for periodic orbits.
\newblock {\em Systems \& control letters}, 26(2):95--105, 1995.

\bibitem{battin1999introduction}
Richard~H Battin.
\newblock {\em An introduction to the mathematics and methods of astrodynamics,
  revised edition}.
\newblock American Institute of Aeronautics and Astronautics, 1999.

\bibitem{belleter2019observer}
Dennis Belleter, Mohamed~Adlene Maghenem, Claudio Paliotta, and Kristin~Y
  Pettersen.
\newblock Observer based path following for underactuated marine vessels in the
  presence of ocean currents: A global approach.
\newblock {\em Automatica}, 100:123--134, 2019.

\bibitem{bishop1975there}
Richard~L Bishop.
\newblock There is more than one way to frame a curve.
\newblock {\em The American Mathematical Monthly}, 82(3):246--251, 1975.

\bibitem{borhaug2010straight}
Even B{\o}rhaug, Alexey Pavlov, Elena Panteley, and Kristin~Y Pettersen.
\newblock Straight line path following for formations of underactuated marine
  surface vessels.
\newblock {\em IEEE transactions on control systems technology},
  19(3):493--506, 2010.

\bibitem{consolini2010path}
Luca Consolini, Manfredi Maggiore, Christopher Nielsen, and Mario Tosques.
\newblock Path following for the pvtol aircraft.
\newblock {\em Automatica}, 46(8):1284--1296, 2010.

\bibitem{de2004astronomia}
Kepler de~Souza Oliveira~Filho and Maria de F{\'a}tima~Oliveira Saraiva.
\newblock Astronomia e astrof{\i}sica.
\newblock {\em Rio Grande do Sul: Livraria da F{\i}sica}, 2004.

\bibitem{fossen2003line}
Thor~I Fossen, Morten Breivik, and Roger Skjetne.
\newblock Line-of-sight path following of underactuated marine craft.
\newblock {\em IFAC proceedings volumes}, 36(21):211--216, 2003.

\bibitem{goldstein2002classical}
Herbert Goldstein, Charles Poole, and John Safko.
\newblock {\em Classical mechanics}.
\newblock Pearson Education Limited, third edition, 2014.

\bibitem{kai2019unified}
J-M Kai, Tarek Hamel, and Claude Samson.
\newblock A unified approach to fixed-wing aircraft path following guidance and
  control.
\newblock {\em Automatica}, 108:108491, 2019.

\bibitem{kapitanyuk2017guiding}
Yuri~A Kapitanyuk, Anton~V Proskurnikov, and Ming Cao.
\newblock A guiding vector-field algorithm for path-following control of
  nonholonomic mobile robots.
\newblock {\em IEEE Transactions on Control Systems Technology},
  26(4):1372--1385, 2017.

\bibitem{kelasidi2017integral}
Eleni Kelasidi, P{\aa}l Liljeb{\"a}ck, Kristin~Y Pettersen, and Jan~Tommy
  Gravdahl.
\newblock Integral line-of-sight guidance for path following control of
  underwater snake robots: Theory and experiments.
\newblock {\em IEEE Transactions on Robotics}, 33(3):610--628, 2017.

\bibitem{marino2011nested}
Riccardo Marino, Stefano Scalzi, and Mariana Netto.
\newblock Nested pid steering control for lane keeping in autonomous vehicles.
\newblock {\em Control Engineering Practice}, 19(12):1459--1467, 2011.

\bibitem{negri2021orbit}
Rodolfo~Batista Negri and FB~de~A Antonio.
\newblock Orbit keeping about the martian moons with a robust path following
  control.
\newblock In {\em 2021 European Control Conference (ECC)}, pages 1675--1680.
  IEEE, 2021.

\bibitem{batista2022autonomous}
Rodolfo~Batista Negri and Ant{\^o}nio~FBA Prado.
\newblock Autonomous and robust orbit-keeping for small-body missions.
\newblock {\em Journal of Guidance, Control, and Dynamics}, 45(3):587--598,
  2022.

\bibitem{negri2022study}
Rodolfo~Batista Negri and Ant{\^o}nio Fernando Bertachini de~Almeida Prado.
\newblock Study on autonomous gravity-assists with a path-following control.
\newblock {\em Advances in the Astronautical Sciences}, 176:1857--1875, 2022.

\bibitem{negri2022autonomous}
Rodolfo~Batista Negri, Ant{\^o}nio Fernando Bertachini de~Almeida Prado, Ronan
  Arraes~Jardim Chagas, and Rodolpho~Vilhena de~Moraes.
\newblock Autonomous rapid exploration in close-proximity of asteroids.
\newblock {\em Journal of Guidance, Control, and Dynamics}, 47(5), 2024.

\bibitem{nelson2007vector}
Derek~R Nelson, D~Blake Barber, Timothy~W McLain, and Randal~W Beard.
\newblock Vector field path following for miniature air vehicles.
\newblock {\em IEEE Transactions on Robotics}, 23(3):519--529, 2007.

\bibitem{nielsen2010path}
Christopher Nielsen, Cameron Fulford, and Manfredi Maggiore.
\newblock Path following using transverse feedback linearization: Application
  to a maglev positioning system.
\newblock {\em Automatica}, 46(3):585--590, 2010.

\bibitem{nielsen2008local}
Christopher Nielsen and Manfredi Maggiore.
\newblock On local transverse feedback linearization.
\newblock {\em SIAM Journal on Control and Optimization}, 47(5):2227--2250,
  2008.

\bibitem{oliveira2016moving}
Tiago Oliveira, A~Pedro Aguiar, and Pedro Encarnacao.
\newblock Moving path following for unmanned aerial vehicles with applications
  to single and multiple target tracking problems.
\newblock {\em IEEE Transactions on Robotics}, 32(5):1062--1078, 2016.

\bibitem{oliveira2013moving}
Tiago Oliveira, Pedro Encarna{\c{c}}ao, and A~Pedro Aguiar.
\newblock Moving path following for autonomous robotic vehicles.
\newblock In {\em 2013 European Control Conference (ECC)}, pages 3320--3325.
  IEEE, 2013.

\bibitem{rubi2020survey}
Bartomeu Rubi, Ramon P{\'e}rez, and Bernardo Morcego.
\newblock A survey of path following control strategies for uavs focused on
  quadrotors.
\newblock {\em Journal of Intelligent \& Robotic Systems}, 98(2):241--265,
  2020.

\bibitem{shtessel2014sliding}
Yuri Shtessel, Christopher Edwards, Leonid Fridman, Arie Levant, et~al.
\newblock {\em Sliding mode control and observation}, volume~10.
\newblock Springer, 2014.

\bibitem{slotine1991applied}
Jean-Jacques~E Slotine, Weiping Li, et~al.
\newblock {\em Applied nonlinear control}, volume 199.
\newblock Prentice hall Englewood Cliffs, NJ, 1991.

\bibitem{stankovic2000decentralized}
Srdjan~S Stankovic, Milorad~J Stanojevic, and Dragoslav~D Siljak.
\newblock Decentralized overlapping control of a platoon of vehicles.
\newblock {\em IEEE Transactions on Control Systems Technology}, 8(5):816--832,
  2000.

\bibitem{sujit2014unmanned}
PB~Sujit, Srikanth Saripalli, and Joao~Borges Sousa.
\newblock Unmanned aerial vehicle path following: A survey and analysis of
  algorithms for fixed-wing unmanned aerial vehicless.
\newblock {\em IEEE Control Systems Magazine}, 34(1):42--59, 2014.

\bibitem{utkin2017sliding}
Vadim Utkin, J{\"u}rgen Guldner, and Jingxin Shi.
\newblock {\em Sliding mode control in electro-mechanical systems}.
\newblock CRC press, 2017.

\bibitem{wang2019cooperative}
Yuanzhe Wang, Danwei Wang, and Senqiang Zhu.
\newblock Cooperative moving path following for multiple fixed-wing unmanned
  aerial vehicles with speed constraints.
\newblock {\em Automatica}, 100:82--89, 2019.

\bibitem{wit1993nonlinear}
C~Canudas~de Wit, Hayate Khennouf, Claude Samson, and Ole~J Sordalen.
\newblock Nonlinear control design for mobile robots.
\newblock In {\em Recent trends in mobile robots}, pages 121--156. World
  Scientific, 1993.

\bibitem{wu20193}
Xinyu Wu, Jia Liu, Chenyang Huang, Meng Su, and Tiantian Xu.
\newblock 3-d path following of helical microswimmers with an adaptive
  orientation compensation model.
\newblock {\em IEEE Transactions on Automation Science and Engineering},
  17(2):823--832, 2019.

\bibitem{yao2020path}
Weijia Yao and Ming Cao.
\newblock Path following control in 3d using a vector field.
\newblock {\em Automatica}, 117:108957, 2020.

\bibitem{zheng2020moving}
Zewei Zheng.
\newblock Moving path following control for a surface vessel with error
  constraint.
\newblock {\em Automatica}, 118:109040, 2020.

\bibitem{zuo2018three}
Zongyu Zuo, Lin Cheng, Xinxin Wang, and Kangwen Sun.
\newblock Three-dimensional path-following backstepping control for an
  underactuated stratospheric airship.
\newblock {\em IEEE Transactions on Aerospace and Electronic Systems},
  55(3):1483--1497, 2018.

\end{thebibliography}



\end{document}